\documentclass[a4paper,11pt]{article}
\usepackage{pos}

\title{Intelligent experiments through real-time AI: Fast Data Processing and Autonomous Detector Control for sPHENIX and future EIC detectors}

 \author*[a]{J. Kvapil}
 \author[b]{G. Borca-Tasciuc}
 \author[c]{H. Bossi}
 \author[d]{K. Chen}
 \author[d]{Y. Chen}
 \author[c]{Y. Corrales Morales}
 \author[a]{H. Da Costa}
 \author[a]{C. Da Silva}
 \author[c]{C. Dean}
 \author[a]{J. M. Durham}
 \author[e]{S. Fu}
 \author[f]{C. Hao}
 \author[c]{P. Harris}
 \author[c]{O. Hen}
 \author[c]{H. Jheng}
 \author[c]{Y. Lee}
 \author[f]{P. Li}
 \author[a]{X. Li}
 \author[a]{Y. Lin}
 \author[a]{M. X. Liu}
 \author[c]{V. Loncar}
 \author[g]{J. P. Mitrevski}
 \author[e]{A. Olvera}
 \author[h]{M. L. Purschke}
 \author[a]{J. S. Renck}
 \author[c]{G. Roland}
 \author[i]{J. Schambach}
 \author[a]{Z. Shi}
 \author[g]{N. Tran}
 \author[j]{N. Wuerfel}
 \author[f]{B. Xu}
 \author[k]{D. Yu}
 \author[f]{H. Zhang}

 \affiliation[a]{Los Alamos National Laboratory,\\Bikini Atoll Rd, Los Alamos, NM 87545, United States}
 \affiliation[b]{Rensselaer Polytechnic Institute,\\ 110 8th St, Troy, NY 12180, United States}
 \affiliation[c]{Massachusetts Institute of Technology,\\77 Massachusetts Ave, Cambridge, MA 02139, United States}
 \affiliation[d]{Central China Normal University,\\ No.152, Luoyu Rd, Wuhan 430079, China}
 \affiliation[e]{University of North Texas,\\1155 Union Cir, Denton, TX 76205, United States}
 \affiliation[f]{Georgia Institute of Technology,\\225 North Ave, Atlanta, GA 30332, United States}
 \affiliation[g]{Fermilab,\\PO Box 500. Batavia IL 60510, United States}
 \affiliation[h]{Brookhaven National Laboratory,\\PO Box 5000 Upton, NY 11973, United States}
 \affiliation[i]{Oak Ridge National Laboratory,\\P.O. Box 2008. Oak Ridge, TN 37831, United States}
 \affiliation[j]{University of Michigan,\\500 S State St, Ann Arbor, MI 481091, United States}
 \affiliation[k]{New Jersey Institute of Technology,\\323 Dr Martin Luther King Jr Blvd, Newark, NJ 07102, United States}

\emailAdd{Jakub.Kvapil@lanl.gov}

\abstract{This R\&D project, initiated by the DOE Nuclear Physics AI-Machine Learning initiative in 2022, leverages AI to address data processing challenges in high-energy nuclear experiments (RHIC, LHC, and future EIC). Our focus is on developing a demonstrator for real-time processing of high-rate data streams from sPHENIX experiment tracking detectors. The limitations of a 15 kHz maximum trigger rate imposed by the calorimeters can be negated by intelligent use of streaming technology in the tracking system. The approach efficiently identifies low momentum rare heavy flavor events in high-rate p+p collisions (3MHz), using Graph Neural Network (GNN) and High Level Synthesis for Machine Learning (hls4ml). Success at sPHENIX promises immediate benefits, minimizing resources and accelerating the heavy-flavor measurements. The approach is transferable to other fields. For the EIC, we develop a DIS-electron tagger using Artificial Intelligence - Machine Learning (AI-ML) algorithms for real-time identification, showcasing the transformative potential of AI and FPGA technologies in high-energy nuclear and particle experiments real-time data processing pipelines.}

\FullConference{%
  42nd International Conference on High Energy Physics (ICHEP2024)\\  
  18-24 July 2024\\
  Prague, Czech Republic
}

\begin{document}
\maketitle

\section{Introduction}

The sPHENIX experiment located at the relativistic heavy ion collider (RHIC) accelerator at Brookhaven National Laboratory is the first experiment with a full calorimetry system at RHIC, with running period of 2022-2025. The sPHENIX physics program consists of cold QCD, jet structure, quarkonium spectroscopy, open heavy flavour, and heavy-flavour jets. The main tracking detectors are 3 layers of silicon pixel detector (MVTX), 2 layers of silicon strip detector (INTT), and gaseous time-projection chamber (TPC). It uses a hybrid data acquisition (DAQ) supporting both triggered and streaming readout. The DAQ is limited by 300 Gb/s bandwidth and due to the sheer volume the TPC produces, only around 10\% of the data can be saved for offline processing.

The main goal of this project is to find beauty-quark signatures by sampling the remaining 90\% of delivered luminosity. The pipeline is as follows, reconstruct tracks from the silicon, search for a displaced vertices based on the decay topology, generate a trigger, and send it to TPC to mark the data to be saved. This allows to sample the remaining 90\% of the luminosity for low $p_\mathrm{T}$ beauty events. As the TPC buffers can hold around 30 $\mathrm{\mu s}$ of data, the stretched goal of this project is to provide a trigger within 10 $\mathrm{\mu s}$. This requires an embedding of machine learning algorithms on the FPGA.

\section{The algorithm pipeline}
The end-to-end pipeline consists of following steps:
\begin{enumerate}
\item Hit decoding and clustering
\item Event building
\item Track construction
\begin{itemize}
\item Edge candidate generation
\item Edge candidate classification
\item Track construction from the edges
\end{itemize}
\item Transverse momentum prediction
\item Trigger detection
\end{enumerate}

The hit clustering and decoding is a conventional logic written for FPGA. The track construction is broken in three pieces. First, the edge candidate generation connects all detector clusters together while applying some geometrical constraints. The detectors clusters are represented as nodes in Graph Neural Network (GNN) and connections between them as edges. It was verified that applying geometrical constraints reduces the number of edges by ~50\% at a small cost of 0.4\% accuracy. Next, a graph convolutional neural network (GCN) \cite{GCN} is used for the classification to predict the true edges. Lastly, the predicted true edges are connected to form a track. 
The next step uses a least-squares method to estimate the curvature of the constructed tracks which is directly proportional to a track momentum. The trigger detection accuracy improved by 15\% while estimating the curvature. The final step is the trigger detection based on Bipartite Graph Network with Set Transformers (BGN-ST) \cite{FastML} that topologically selects a heavy flavour signal. The BGN-ST is an attention-based algorithm that allows modeling of the following information
\begin{enumerate}
    \item Local track-to-track interactions that determine whether two tracks share the same vertex
    \item Track-to-global interactions that determine the collision vertex
    \item Global-to-track interactions that determine whether the track origin vertex is centered around the collision vertex
\end{enumerate}
The algorithm takes the initial track information, assigns weights to each link, assigns them to a particular vertex, and using feed-forward neural network updates the track information to repeat the process. There are a total of 37 features in the track node vector, for example: 5 clusters (3 MVTX + 2 INTT), the length between each clusters (edges), the angle between edges, the total length of edges, and the track radius (momentum). The BGN-ST architecture main element is called SEBA (Set Attention and Bipartite Aggregation). The model was generated using PyTorch and PyTorch Geometric and initial training was done on simulated data from the MVTX and INTT with sPHENIX geometry.

\section{Model accuracy studies}
\subsection{Charm decays}
The first case of study was done on $\mathrm{D^0\rightarrow K\pi}$ meson decays. All the accuracies reported in this document were calculated by using 50\% signal-to-background ratio data samples. The first version of BGN-ST model compared to other baseline modesl is shown in Table \ref{tab:baseline-r}. The BGN-ST provides the highest accuracy and estimating the momentum $p_\mathrm{T}$ via the track curvature increases the accuracy by 13.43\%. A second version of the model is currently available with two major improvements. First, a data augmentation which implements robustness against detector alignment variation by perturbating hits to a different point on particle trajectory. Second, an additional adjacency matrix was added to a loss function to account for whether two tracks come from the same parent. The improved model version increased accuracy from 87.56\% to 90.22\% while selecting the $\mathrm{D^0}$ signal. This translates to 23.2\% efficiency and 2.3\% purity for a 0.1\% signal/background sample with 99\% background rejection rate. This is a 2.3 fold improvement in efficiency from the current sPHENIX standard and 23x rate increase for the tagging purity compared to random selection. Final study for $\mathrm{D^0}$ decays involved how it compares to a simpler model where the trigger detection is based on clustered hits in layers instead of reconstructing tracks. The hit-based model implemented using GCN provided a 85\% accuracy, 5.22\% less than model with the track construction.

\begin{table}[h]
\caption{Accuracy and Area under the receiver-operating characteristic curve (AUC) of BGN-ST model triggering on $D^0$ decays, comparing to other baseline models, and whether a track radius was estimated. Older version of the table can found at \cite{FastML}.}
\label{tab:baseline-r}
  \centering
\begin{tabular}{c c c c c c c c}
 \hline
& \multicolumn{3}{c}{with track radius estimation} & \phantom{abc} &  \multicolumn{3}{c}{without track radius estimation} \\
Model & \# Parameters & Accuracy & AUC && \# Parameters & Accuracy & AUC\\
\hline
Set Transformer & 299,266 & 86.40\% & 91.92\% && 298,882 & 72.04\% & 78.92\% \\
GarNet & 284,210 & 86.22\% & 91.81\% && 284,066 & 72.59\% & 79.61\%  \\
PN+SAGPool & 780,934 & 86.25\% & 92.91\% && 780,678 & 69.22\% & 77.18\%\\
BGN-ST & 363,426 & \textbf{87.56\%} & \textbf{93.22\%} && 363.170 & \textbf{74.13\%} & \textbf{81.81\%} \\
\hline
\end{tabular}
\end{table}

\subsection{Beauty decays}
The production of $\mathrm{D^0}$ is plentiful at sPHENIX and any additional trigger would exceed the DAQ capability, thus the main focus of this project is on beauty decays with have around 0.05\% production probability at RHIC energies. The BGN-ST model reached 97.38\% accuracy in detecting beauty decays. Implementing trigger detection directly from the hits using GarNet \cite{GarNet} provided 90.57\% accuracy and trigger detection from edge candidates using Graph Attention Network \cite{GraphAtt} provided 91.57\% accuracy. Thus, reconstructing the tracks provides a significant boost in accuracy compared to cluster-based algorithms. The attention network provided a slight improvement in accuracy. The last test of the hit-based GarNet model was to estimate the robustness against noise. For around 350 physics-produced hits 65 random noise hits were added, which corresponds to the noise level of $10^{-7}$, an exaggerated noise level from expected value of $10^{-9}$. The cluster detection pipeline based on GarNet provided 88.52\% accuracy with the noise present, a small decrease in accuracy. A purity and efficiency is currently under investigation.

\section{Generation of FPGA code}
The models need to be translated from Python into VHDL/Verilog code that is used for deployment on the FPGA. We are implementing two approaches, the first one is a manual rewrite of the code into C++ and using FlowGNN \cite{cit:flowgnn} to generate VHDL code, the second is using automated translation based on hls4ml \cite{cit:hls}. The FELIX-712 board, which contains a Xilinx Kintex UltraScale FPGA, was chosen as a target device for following reasons: it contains enough optical links to receive the required data, it has large firmware and software support, and it is already in use at sPHENIX to read out MVTX, INTT, and TPC. 

\paragraph{Track-based model (BGN-ST)} The first implementation of the edge candidates classification from the track-based model resulted in 150 $\mathrm{\mu s}$ latency at 130 MHz clock which is much over the desired target. A further iteration with the model developers to lighten the model led into a second version of the edge classification that offered 8.82 $\mathrm{\mu s}$ latency at 285 MHz clock. The resource utilization on a Alveo U280 aggregator card (approximately twice as big as FELIX-712) was following: 194K (14.9\%) Lookup tables (LUT), 214K (8.2\%) Flip Flops (FF), 406 (20.2\%) Block Random Access Memory (BRAM), and 488 (5.4\%) Digital Signal Processor (DSP). 

The support of PyTorch Geometric used in the BGS-ST model is currently not supported in hls4ml. The developers expect to release support in 2025, it is expected that the edge candidate classification can reach nanosecond latencies.

\paragraph{Hit-based model (GarNet)}As translating another parts of the model would be very labor intensive until the hls4ml support is provided, a current focus shifted to an implementation of hit-based beauty model in GarNet. The translated hit-based model in FlowGNN offered a start-to-end latency of 9.2~$\mathrm{\mu s}$ at 180 MHz clock, due to element reuse, this is also the trigger dead-time. The current focus of the team is to do a detailed latency breakdown based on the FlowGNN parameters and explore a parallelism options. 

The first translation using hls4ml provided fully pipelined design with latency 505 ns and following utilisation: 42K LUT (6.32\%), 43.2K FF (3.25\%), 311 BRAM (1.41\%), and 574 DSP (10.4\%). A current focus of the team is to optimise the utilisation with respect to the latency.

\section{Summary}

An embedding of machine learning algorithms on FPGA's can provide a huge improvement in online reconstruction and event filtering at high energy physics experiments. Our attention-based model provides the best accuracy while comparing to other models, and a significant improvement in beauty event selection at sPHENIX. As of today, all the pieces of the firmware with simplified hit-based model are written, validated, and are being combined on a single FPGA, following an optimisation stage to achieve the desired utilisation and latency. Testing with the full system is expected by the end of the year, followed by the translation of attention-based model using updated hls4ml software in 2025. The next focus would be in designing the model for future EIC experiments.

\end{document}